\documentclass[
aps,%
12pt,%
final,%
notitlepage,%
oneside,%
onecolumn,%
nobibnotes,%
nofootinbib,
superscriptaddress,%
noshowpacs,%
centertags]%
{revtex4}

\usepackage{slashed}
\usepackage{natbib}

\RequirePackage{amssymb,amsmath}
\RequirePackage{eufrak}
\RequirePackage{mathtext}
\RequirePackage[cp866]{inputenc}
\RequirePackage[T2A]{fontenc}
\RequirePackage[russian,english]{babel}

\RequirePackage{bm}
\RequirePackage{array}
\RequirePackage{longtable}
\RequirePackage{graphicx}
\RequirePackage{calc}
\RequirePackage{ifthen}

\RequirePackage{caption2}
\setcaptionwidth{\linewidth}

\renewcommand{\captionfont}{\footnotesize}
\newcaptionstyle{nonumber}{%
  \usecaptionmargin%
  \captionfont%
  \onelinecaption%
    {{\parbox[t]{\realcaptionwidth}{\captiontext}}}%
    {\captiontext}
}%

\makeatletter
\renewcommand \thefigure{\@arabic\c@figure}
\renewcommand \thetable{\@arabic\c@table}
\def\place@bibnumber@inl#1{#1.}%

\def\thesection       {\arabic{section}}
\def\p@section        {}
\def\thesubsection    {\thesection.\arabic{subsection}}
\def\p@subsection     {\thesection.}

\def\p@subsubsection  {\thesection\,\thesubsection\,}

\def\section{%
  \@startsection
    {section}%
    {1}%
    {\z@}%
  {0.8cm \@plus1ex \@minus .2ex}%
  {0.5cm \@plus1ex \@minus.2ex}%
    {%
      \normalfont
      \centering
      \MakeTextUppercase
    }%
}%
\def\@sectioncntformat#1{\csname the#1\endcsname.\quad}%
\def\subsection{%
  \@startsection
    {subsection}%
    {2}%
    {\z@}%
    {.8cm \@plus1ex \@minus .2ex}%
    {.5cm \@plus1ex \@minus.2ex}%
    {%
     \normalfont\slshape
     \centering
    }%
}%
\def\subsubsection{%
  \@startsection
    {subsubsection}%
    {3}%
    {\z@}%
    {.8cm \@plus1ex \@minus .2ex}%
    {.5cm \@plus1ex \@minus.2ex}%
    {%
     \normalfont\small\itshape
     \centering
    }%
}%
\def\refitem#1{\ifthenelse{\boolean{hmode}}{\protect\newline{}}{}%
\textbackslash{}refitem\{\hbox{\bfseries{}%
\itshape{}#1}\}\ifthenelse{\boolean{hmode}}{\protect\linebreak{}}{}}%
\makeatother
\def\refitem#1{\relax}

\begin{document}
\selectlanguage{english} 
\title{Associated quarkonia production in a single boson \\ $e^+e^-$~annihilation}
%
\author{
\firstname{I.~N.} 
\surname{Belov}}
\affiliation{INFN, Sezione di Genova, Italy}
\author{
\firstname{A.~V.} 
\surname{Berezhnoy}}
\affiliation{SINP MSU, Moscow, Russia}
\author{
\firstname{E.~A.} 
\surname{Leshchenko}}
\affiliation{Physics department of MSU, Moscow, Russia}
\begin{abstract}
The production cross sections of charmonia, charmonium-bottomonium and bottomonia pairs in a single boson $e^+e^-$~annihilation have been studied in a wide range of energies, which will be achieved at future $e^+e^-$~colliders such as ILC and FCC. 
One loop QCD corrections to QCD and EW contributions as well as their interference are considered. 
The both intermediate bosons $\gamma$ and $Z$ are taken into account. 
\end{abstract}
\maketitle
\section{Introduction}
Heavy quark physics have been remaining attractive for theorists and experimentalists throughout its long history. 
Nearly every year is now marked with discoveries in this field as a result of various experiments such as LHC, BELLE-II and the BES-III.
The production of quarkonium pairs is a popular topic of discussions.
One of the most intriguing researches is the observation of $J/\psi \, \eta_c$ pairs in the $e^+e^-$~annihilation where the experimental yield measured at BELLE and BaBar~\cite{Abe:2004ww,Aubert:2005tj} was underestimated by the theoretical predictions~\cite{Braaten:2002fi} by the order of magnitude. 
This event prompted the countless investigations~
\cite{Dong:2012xx,Xi-Huai:2014iaa,Feng:2019zmt,Zhang:2005cha,Gong:2007db,
Bondar:2004sv,Braguta:2005kr,Berezhnoy:2006mz,Braguta:2006nf,Bodwin:2006dm,Ebert:2006xq,Berezhnoy:2008zz,Ebert:2008kj,Braguta:2008hs,Braguta:2008tg,Sun:2009zk,Braguta:2012zza,Sun:2018rgx}, the results of which led to a decent level of data agreement.
Another surge of interest to this topic occurred in 2020, when the LHCb Collaboration published article~\cite{Aaij:2020fnh} about the observation of the structure in the $J/\psi~J/\psi$ spectrum at large statistics.
 
All these results have motivated us to study the processes of paired quarkonium production, specifically the production of $J/\psi \, \eta_c$ and $J/\psi \, J/\psi$ pairs and also $\Upsilon \, \eta_b$ and $\Upsilon \, \Upsilon$ pairs in the $e^+e^-$~annihilation.
Proper observation of such processes is still impossible in the currently existing experiments due to low achieved collision energy.
Nevertheless, current work may be of practical interest in terms of several discussed future projects, such as ILC and FCC with announced energy ranges as high as $\sqrt{s}~=~90\div400~\text{GeV}$~\cite{Koratzinos:2014cla} and $\sqrt{s}~=~250~\text{GeV}$~\cite{KEKInternationalWorkingGroup:2019spu} correspondingly, aimed to investigations at the energies of the order of $Z$-boson's mass.
As it was emphasised in the Conceptual Design Report~\cite{Abada:2019}, it is planned to obtain as much as $5\times10^{12}$ decays of $Z$~bosons, with energies of about $\sqrt{s}~\approx~91~\text{GeV}$, which will result into the outstanding luminosity on the facility.
And such enormous amount of statistics to be obtained on FCC makes it one of the most perspective successor of B-factories for heavy hadron research field, including charmonia and bottomonia investigations in a wide energy range.
We also consider studied processes interesting in terms of the $Z$-boson decays to the charmonia and the bottomonia, which may have a certain potential for the experiments at the LHC, see~\cite{Sirunyan:2019lhe}.

Our previous studies involved the investigation of the $B_c$ pair production~\cite{Berezhnoy:2016etd}, the $J/\psi \, J/\psi$ and the $J/\psi \, \eta_c$ pair production~\cite{Berezhnoy:2021tqb} around the $Z$ mass within the NLO approximation considering only QCD contribution, as well as the production of $J/\psi \, \eta_b$ and $\Upsilon \, \eta_c$ pairs considering QCD contribution within NLO approximation and EW contribution within LO accuracy~\cite{Lesh:2021}.
The substantial conclusion from the listed researches is that the loop corrections for investigated processes essentially affect the cross section values.
This result is consistent with findings by other research groups investigating the paired quarkonium production in the $e^+e^-$annihilation.
As an extension of this study  we investigate the electroweak contribution in NLO approximation to these processes and implement the derived calculation technique for bottomonia case: the $\Upsilon \, \Upsilon$ and the $\Upsilon \, \eta_b$ pair production.

Thus, the following processes are studied in this paper: ${e^{+}e^{-}\xrightarrow{\gamma^*,\ Z^*}~J/\psi \, \eta_c}$, ${e^{+}e^{-}\xrightarrow{ Z^* }~J/\psi \, J/\psi}$, ${e^{+}e^{-}\xrightarrow{\gamma^*,\ Z^*}~\Upsilon \, \eta_b}$ and ${e^{+}e^{-}\xrightarrow{ Z^*}~\Upsilon \, \Upsilon}$.


\section{Methods}

The production features of quarkonia pairs in a single boson $e^+e^-$~annihilation are determined by the certain set of selection rules:
\begin{itemize}
    \item The productions of both vector-vector pairs (VV) and pseudoscalar-pseudoscalar (PP) pairs through the intermediate photon and the vector part of $Z$ vertex are prohibited due to the charge parity conservation. 
    \item The productions of vector-pseudoscalar (VP) pairs via the axial part of $Z$ vertex is prohibited for the very same reason. 
    \item The PP pairs production  via the axial part of $Z$ vertex is prohibited due to the combined $CP$ parity conservation.
\end{itemize}

These selection rules acted as additional verification criteria of calculations.

There are two production mechanisms for the investigated processes.
The first one is the single gluon exchange, for which the  tree level contribution is ${\cal O}(\alpha^2 \alpha_s^2)$. 
The second production mechanism is the single photon or $Z$ boson exchange.
In that case the tree level contribution is of order ${\cal O}(\alpha^4)$. 
In text we refer to this contributions as QCD LO and EW LO contributions correspondingly.

We also take into account QCD one-loop correction to both QCD LO and EW LO contributions, which are of orders ${\cal O}(\alpha^2 \alpha_s^3)$ and ${\cal O}(\alpha^4 \alpha_s)$ correspondingly.  We refer to them as QCD NLO and EW NLO contributions.

Thus, when studying these processes, one should take into account 7 contributions to the total cross sections: 
\begin{multline}
    |\mathcal{A}|^{2} = |\mathcal{A}^{LO}_{QCD}|^2 + |\mathcal{A}^{LO}_{EW}|^2 + 2Re(\mathcal{A}^{LO}_{QCD}\mathcal{A}^{LO*}_{EW}) + \\ 
    2Re(\mathcal{A}^{NLO}_{QCD}\mathcal{A}^{LO*}_{QCD}) + 2Re(\mathcal{A}^{NLO}_{QCD}\mathcal{A}^{LO*}_{EW}) + \\
    2Re(\mathcal{A}^{NLO}_{EW}\mathcal{A}^{LO*}_{QCD}) + 2Re(\mathcal{A}^{NLO}_{EW}\mathcal{A}^{LO*}_{EW}) + \dots \ .
\end{multline}

To describe the  double heavy quarkonia the Nonrelativistic QCD (NRQCD)~\cite{Bodwin:1994jh} is used.
The NRQCD is based on the hierarchy of scales for the quarkonia: $m_q >> m_q v, m_q v^2, \Lambda_{QCD}$, where $m_q$ is the mass of the heavy quark and $v$ is  the heavy quark  velocity in the quarkonium. 
Such formalism allows to divide the investigated process into the hard subprocess of heavy quarks production and the soft fusion of heavy quarks into  quarkonia. 



In order to construct the bound states we  put $v=0$  and replace the spinor products $v(p_{\bar{q}})\bar{u}(p_q)$ by the appropriate covariant projectors for color-singlet spin-singlet and spin-triplet states:
\begin{align}
    &\Pi_{P}(Q_{q},m_{P})=\frac{\slashed Q-2 m_{P}}{2\sqrt{2 }}\gamma^{5}\otimes \frac{\boldsymbol 1}{\sqrt{N_c}},
    &\Pi_{V}(P_{q},m_{V})=\frac{\slashed P-2 m_{V}}{2\sqrt{2}}\ \slashed \epsilon^{V} \otimes \frac{\boldsymbol 1}{\sqrt{N_c}}, 
\end{align}
where $Q_q$, $m_{P}$ and $P_q$, $m_{V}$ are momenta and masses of the pseudoscalar and vector final states correspondingly. 
Polarization $\epsilon^{V}$ of the vector meson satisfy the following constraints: $\epsilon^{V}~\cdot~{\epsilon^{V}}^*~=~-1$, $\epsilon^{V}~\cdot~P_q~=~0$. 
These operators enclose the fermion lines into traces.

The renormalization procedure must be applied to the one-loop contribution. The so-called \glqq On-shell\grqq~scheme has been used for renormalization of masses and spinors and $\overline{MS}$ scheme has been adopted for coupling constant renormalization:
\begin{align}
    Z_{m}^{OS}=1-\frac{\alpha_{s}}{4\pi}C_{F}C_{\epsilon}\left[\frac{3}{\epsilon_{UV}}+4\right]+\mathcal{O}\left(\alpha_{s}^2\right), \nonumber \\ 
    Z_{2}^{OS}=1-\frac{\alpha_{s}}{4\pi}C_{F}C_{\epsilon}\left[\frac{1}{\epsilon_{UV}}+\frac{2}{\epsilon_{IR}}+4\right]+\mathcal{O}\left(\alpha_{s}^2\right),\\
    Z_{g}^{\overline{MS}}=1-\frac{\beta_{0}}{2}\frac{\alpha_{s}}{4\pi}\left[\frac{1}{\epsilon_{UV}}-\gamma_{E}+\ln{4\pi}\right]+\mathcal{O}\left(\alpha_{s}^2\right),\nonumber
\end{align}
where $C_\epsilon = \left(\frac{4\pi\mu^2}{m^2}e^{-\gamma_E}\right)^\epsilon$ and $\gamma_E$ is the Euler constant.




The counter-terms are obtained from the leading order diagrams.
The isolated singularities are then cancelled with the singular parts of the calculated counter-terms.

The \texttt{FeynArts}-package~\cite{Hahn:2000kx} in Wolfram Mathematica is used to generate the diagrams and the accompanying analytical amplitudes.

In total  6 nonzero tree level EW, 10 nonzero one loop EW and 4 nonzero tree level QCD diagrams contribute to PV pair production ( $e^+e^-~\xrightarrow{\gamma^*,Z^*}~J/\psi \, \eta_c$ and $e^+e^-~\xrightarrow{\gamma^*,Z^*}~\Upsilon \, \eta_b$). 
The number of nonzero one-loop QCD diagrams depends on the intermediate boson type: 80 in case of virtual photon and 92 in case of virtual $Z$ boson.
The VV pair production subprocesses $e^+e^-~\xrightarrow{Z^*}~J/\psi \, J/\psi$ and $e^+e^-~\xrightarrow{Z^*}~\Upsilon \, \Upsilon$ are described by 8 nonzero tree level EW diagrams, 4 nonzero tree QCD diagrams, 20 nonzero one loop EW diagrams and 86 nonzero one loop QCD diagrams.  



To calculate the tree level amplitudes we use  \texttt{FeynArts}~\cite{Hahn:2000kx} and \texttt{FeynCalc}~\cite{Shtabovenko:2020gxv} packages in \texttt{Wolfram Mathematica}, while the computation of loop amplitudes demands the following toolchain:
$\texttt{FeynArts}\rightarrow\texttt{FeynCalc}(\texttt{TIDL})\xrightarrow{}\texttt{Apart~\cite{Feng:2012iq}}\rightarrow\texttt{FIRE~\cite{Smirnov:2008iw}}\rightarrow\texttt{X~\cite{Patel:2016fam}}$.

The \texttt{FeynCalc} package performs all required algebraic operations with Dirac and color matrices, in particular the trace evaluation.
The  Passarino-Veltman reduction is carried out with the implementation of \texttt{TIDL} library included in \texttt{FeynCalc} package.
The \texttt{Apart} function performs the additional simplification providing the partial fractioning of IR-divergent integrals.
The integrals produced in the above described phases are completely reduced to master integrals by the FIRE package.
At last, by using the \texttt{X}-package the master integrals are evaluated by substituting their analytical expressions.

The NLO amplitudes are computed using the conventional dimensional regularization (CDR) approach with a $D$-dimensional loop and external momenta.

The so-called \glqq naive\grqq~prescription  for $\gamma^5$ was implemented: $\gamma^5$ anticommutes with all other $\gamma$ matrices,
the remaining  $\gamma^5$ in traces with an odd number of $\gamma^5$ matrices is shifted to the right  and then replaced by
\begin{equation*}
    \gamma^{5}=-\frac{i}{24}\varepsilon_{\alpha\beta\sigma\rho}\gamma^{\alpha}\gamma^{\beta}\gamma^{\sigma}\gamma^{\rho}.
\end{equation*}
 
The strong coupling constant was treated within the two loops accuracy:
\begin{equation*}
    \alpha_S (\mu) = \frac{4\pi}{\beta_0 L}\left(1-\frac{\beta_1 \ln{L}}{\beta_0^2 L}\right),
\end{equation*}
where $L = \ln{\mu^2/\Lambda^2}$, $\beta_0 = 11-\frac{2}{3} N_f$,  $\beta_1 = 10-\frac{38}{3} N_f$, 
and $\alpha_S (M_Z) = 0.1179$. 
The renormalization and coupling constant scales are set to be equal:  $\mu_R = \mu$. 
The fine structure constant is fixed at the Thomson limit: $\alpha = 1/137$. 
$u$-, $d$- and $s$-quarks are considered massless. 
The numerical values of other parameters are outlined in Table~\ref{Table:1}.

\section{Results}

\subsection*{Analytical results}

In this subsection we present some analytical results obtained for the discussed  the amplitudes and cross sections.

The asymptotic behaviour of the  amplitudes are presented below:
\begin{align}
    \mathcal{A}^{LO}_{QCD}\sim\frac{1}{s^2};& \ \ \ \ \ 
    \frac{\mathcal{A}^{NLO}_{QCD}}{\mathcal{A}^{LO}_{QCD}} \sim \alpha_s \left( c^{QCD}_2 \ln^2 \left(\frac{s}{m^2_q}\right) + c^{QCD}_1 \ln \left(\frac{s}{m^2_q}\right)+ c^{QCD}_0 + c^{QCD}_\mu \ln \left(\frac{\mu}{m_q}\right) \right); \label{eq:asym_QCD}\\
    \mathcal{A}^{LO}_{EW}\sim\frac{1}{s};&  \ \ \ \ \ 
    \frac{\mathcal{A}^{NLO}_{EW}}{\mathcal{A}^{LO}_{EW}} \sim \alpha_s \left( c^{EW}_1 \ln \left(\frac{s}{m^2_q}\right)+ c^{EW}_0 \right),  \label{eq:asym_EW}  
\end{align}
where $c^{QCD}_i$ and $c^{EW}_i$ are  coefficients independent of the collision energy, the quark masses and the chosen scale.

Since the tree-level amplitudes for the investigated processes have a trivial  Lorentz structure, the analytical expressions for the appropriate cross sections  are  quite simple:

\begin{align}
\sigma^{LO}_{QCD}\left(V P\right) &= 
\frac{131072 \ \pi^3 \alpha ^2 \alpha_s^2 e_q^2 {\cal O}_{V} {\cal O}_{P} \left(s-16 m_q^2\right)^{3/2}}{243 \ s^{11/2}} \
\left( 1+q_{\gamma Z}+q_{Z} \right),\label{eq:LOQCDVP} \\
\sigma^{LO}_{EW}\left(V P\right) &= 
\frac{32 \ \pi^3 \alpha^4 e_q^6 {\cal O}_{V} {\cal O}_{P} \left(s-16 m_q^2\right)^{3/2} \left(s+\frac{16}{3} m_q^2\right)^2}{3 \ m_q^4 s^{11/2}} 
\left( 1+q_{\gamma Z}+q_{Z}\right), \label{eq:LOEWVP}\\
\sigma^{LO}_{QCD}\left(V_1 V_2\right) &= 
\frac{1}{2!}
\frac{512 \ \pi^3 \alpha^2 \alpha_s^2 {\cal O}_{V}^2 \left(s-16 m_q^2\right)^{5/2} }{243 \ s^{9/2}}
\frac{\sec^4{\theta_w}\left(\csc^4{\theta_w}-4\csc^2{\theta_w}+8\right)}{\left(M_Z^2-s\right)^2+\Gamma^2 M_Z^2},  \\
\sigma^{LO}_{EW}\left(V_1 V_2\right) &= 
\frac{1}{2!}
\frac{\pi^3 \alpha^4 e_q^4 {\cal O}_{V}^2 \left(s-16 m_q^2\right)^{5/2} \left(s+\frac{8}{3} m_q^2\right)^2}{6 \ m_q^4 s^{9/2} }
\frac{\sec^4{\theta_w}\left(\csc^4{\theta_w}-4\csc^2{\theta_w}+8\right)}{\left(M_Z^2-s\right)^2+\Gamma^2 M_Z^2},
\end{align}
where $q_{\gamma Z}$ in \eqref{eq:LOQCDVP} and \eqref{eq:LOEWVP} corresponds to the interference of photonic and $Z$~bosonic amplitudes, $q_{Z}$ corresponds to the $Z$~boson annihilation amplitude squared, and $1$ stands for the photon annihilation amplitude squared:
\begin{align}
q_{\gamma Z}&=\frac{\tan^2{\theta_w} \left(\csc^2{\theta_w}-4\right) \left(\csc^2{\theta_w}-4 |e_q|\right)}{8 \ |e_q|}\ 
\frac{ s\left(s-M_Z^2\right)}{ (M_Z^2-s)^2 + \Gamma^2 M_Z^2}, 
\\
q_{Z}&=\frac{\tan^4{\theta_w} \left(\csc^4{\theta_w}-4 \csc^2{\theta_w}+8\right) \left(\csc^2{\theta_w}-4 |e_q|\right)^2}{128 \ e_q^2}\ 
\frac{s^2}{ \left(M_Z^2-s\right)^2 + \Gamma^2 M_Z^2}. 
\end{align}

The analytical expressions for the one-loop cross sections are  cumbersome, and this is why we do not present them in the text. 


\subsection*{Numerical results}

The numerical results  are encapsulated in the Table \ref{Table:2}.
 
\begin{figure}[ht]
    \setcaptionmargin{5mm}
    \onelinecaptionsfalse
    \includegraphics[width = \linewidth]{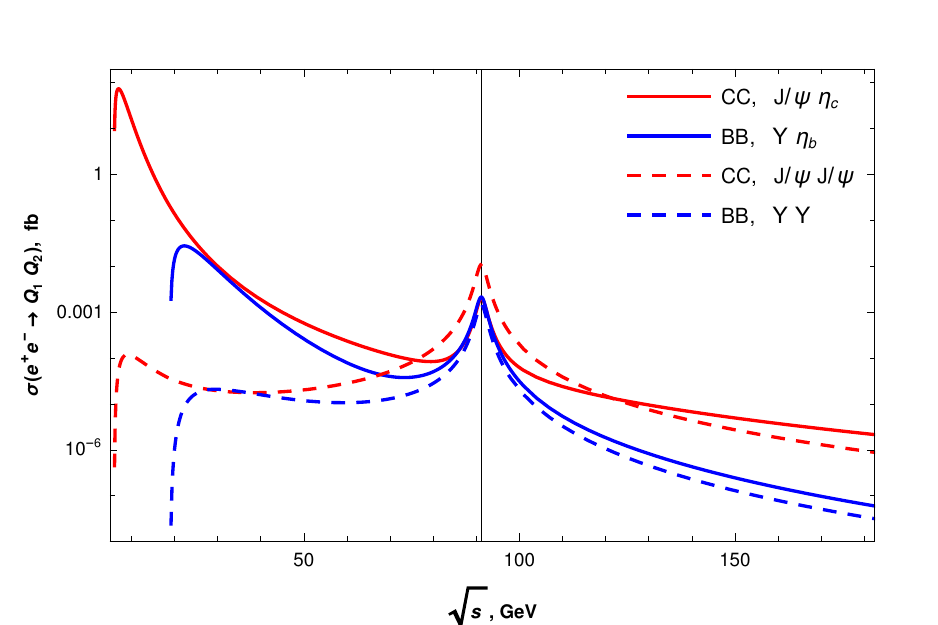}
    \captionstyle{normal}
    \caption{ The total cross sections dependence on the collision energy for the double charmonia and the double bottomonia production.}
    \label{Figure:1}
\end{figure}

As it is seen in Fig.~\ref{Figure:1},   $J/\psi \, \eta_c$  pairs production smaller  than $J/\psi \, J/\psi$ only  near the $Z$-pole, while  $\Upsilon \, \eta_b$ pairs production greater than  $\Upsilon \, \Upsilon$  at all investigated energies.

\begin{figure}[ht]
    \setcaptionmargin{5mm}
    \onelinecaptionsfalse
    $\begin{array}{cc}
        \includegraphics[page = 1, width = 0.5\linewidth]{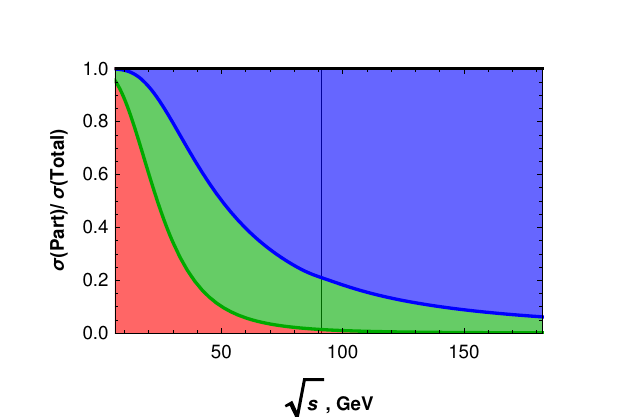} &
        \includegraphics[page = 2, width = 0.5\linewidth]{pictures/sigma/Fig2.pdf} \\
        (a)~J/\psi \, \eta_c & (b)~J/\psi \, J/\psi \\
        \includegraphics[page = 3, width = 0.5\linewidth]{pictures/sigma/Fig2.pdf} &
        \includegraphics[page = 4, width = 0.5\linewidth]{pictures/sigma/Fig2.pdf} \\
        (c)~\Upsilon \, \eta_b & (d)~\Upsilon \, \Upsilon 
    \end{array}$
    \captionstyle{normal}
    \caption{ Relative contributions of different different mechanisms to the total cross section values  as a function of collision energy: the QCD mechanism (red),
    the EW mechanism (blue), the interference of QCD and EW production mechanisms (green).}
    \label{Figure:2}
\end{figure}

The relative QCD and EW contributions to the total cross sections are shown in Fig.~\ref{Figure:2}.
It is worth to note that in the charmonia production the QCD contribution is dominant up to energies of $\sim25~\text{GeV}$, while at higher energies the EW contribution dominates.
As for the associated bottomonia production, at low energies the QCD mechanism appears to be dominant, while at energies above the $Z$-boson threshold both contributions are comparable.

\subsection*{K-factors}

In order to estimate the contribution of NLO corrections to the cross section the  K-factors were numerically calculated for each of the production mechanisms. 

\begin{figure}[ht]
    \setcaptionmargin{5mm}
    \onelinecaptionsfalse
    \includegraphics[width = \linewidth]{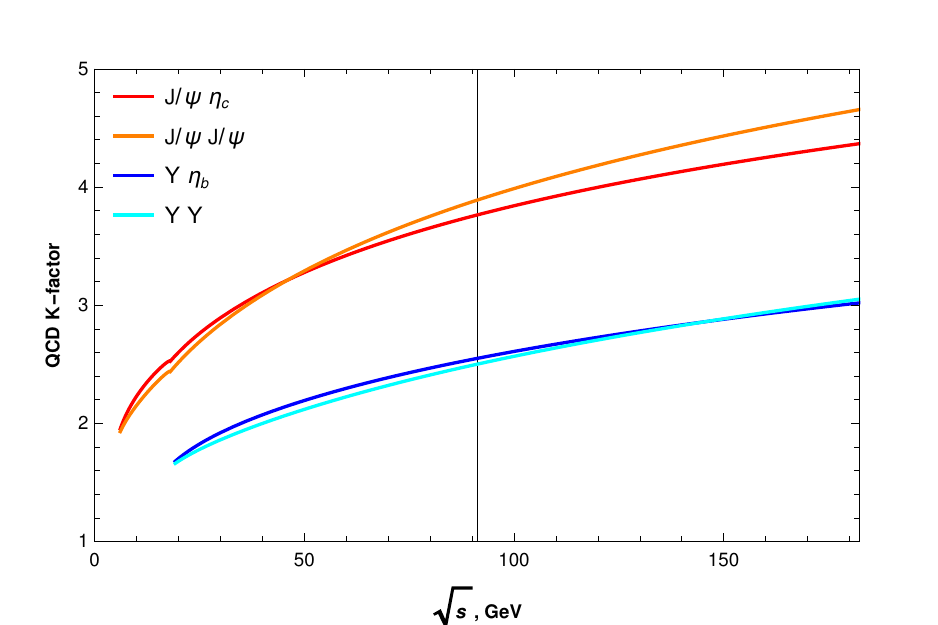} 
    \captionstyle{normal}
    \caption{The total K-factor dependence on the collision energy in case of QCD production.}
    \label{Figure:3}
\end{figure}

\begin{figure}[ht]
    \setcaptionmargin{5mm}
    \onelinecaptionsfalse
    \includegraphics[width = \linewidth]{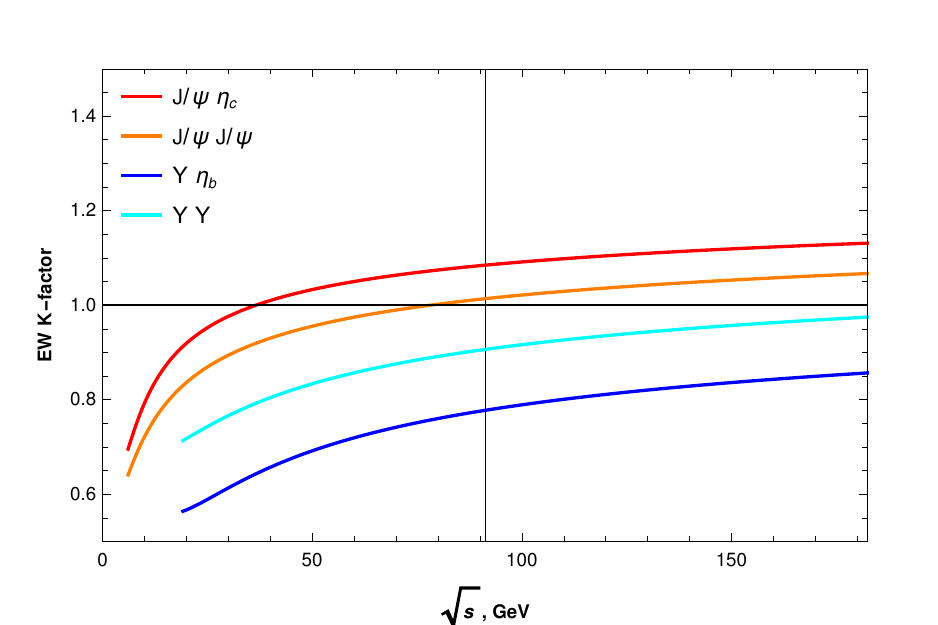}
    \captionstyle{normal}
    \caption{The total K-factor dependence on the collision energy in case of EW production.}
    \label{Figure:4}
\end{figure}

The K-factors dependencies on interaction energy are shown in Fig.~\ref{Figure:3} and Fig.~\ref{Figure:4}. All  K-factors increase with  energy as expected from \eqref{eq:asym_QCD} and \eqref{eq:asym_EW}. 
The  K-factors for the QCD mechanism  are consimilar for PV and VV pairs production, whereas  K-factors for the EW mechanism are sufficiently depend on quarkonia quantum numbers. 
The NLO corrections to QCD mechanism  contribute positively  to the cross section at all investigated energies.
The NLO corrections to the EW charmonia production are negative  at low energies and positive at high energies. The NLO corrections to the EW bottomonia  production are negative at all energies.

The obtained results highlight the significance of NLO corrections to the both discussed mechanisms. 

\section{Conclusions}

The exclusive production of  charmonia and bottomonia pairs  ($J/\psi \, \eta_c$, $J/\psi \, J/\psi$, $\Upsilon \, \eta_b$ and $\Upsilon \, \Upsilon$) in a single boson $e^+e^-$~annihilation has been investigated at interaction energies from the production threshold to $2M_Z$  within the QCD NLO accuracy.

It has been shown that the both studied mechanisms, namely QCD and EW ones, significantly  contribute to the  cross section in a wide energy range. Analytical expressions for QCD and EW amplitudes and cross sections  have been obtained and numerical results for cross sections have been presented. 
Also, the relative contributions of production mechanisms to the total cross section have been analysed. We believe that the presented results may be of considerable interest for experiments at future $e^+e^-$~colliders.

Authors are grateful to the organizing committee of ICPPA2022 for the opportunity to make this report.
This study was conducted within the scientific program of the National Center for Physics and Mathematics, section \#5 \glqq Particle Physics and Cosmology\grqq. Stage 2023-2025.
I.~Belov acknowledges the support from \glqq BASIS\grqq~Foundation, grant No. 20-2-2-2-1.

%
\newpage
%
\newpage
%
\newpage
%
\newpage
%
\newpage
%
\begin{table}[ht] 
    \setcaptionmargin{0mm} 
    \onelinecaptionstrue
    \captionstyle{flushleft}
    \caption{ The cross section values in fb units at different collision energies.}
    \label{Table:1}
    \bigskip
    \begin{tabular}{cccc}
        \hline
        $m_c$ = 1.5~GeV~~  & ~~$m_b$ = 4.7~GeV~~  & ~~$M_Z$ = 91.2~GeV~~  & ~~$\Gamma_Z$ = 2.5~GeV \\
        \hline
    \end{tabular}
    \begin{tabular}{ccc}
        $\langle O \rangle_{J/\psi} =\langle O \rangle_{\eta_c} = 0.523\mbox{ GeV}^3$ & $\langle O \rangle_{\Upsilon} = \langle O \rangle_{\eta_b} = 2.797\mbox{ GeV}^3$ & $\sin^2\theta_w$ = 0.23 \, \\
        \hline
    \end{tabular}
\end{table}
\newpage
\begin{table}[ht] 
    \setcaptionmargin{0mm} 
    \onelinecaptionstrue
    \captionstyle{flushleft}
    \caption{ The cross section values in fb units at different collision energies.}
    \label{Table:2}
    \bigskip
    \begin{tabular}{|c||c|c|c|c|c|c|}
        \hline
        \textbf{E, GeV} & \textbf{15} & \textbf{20} & \textbf{30} & \textbf{50} & \textbf{90} & \textbf{180}  \\
        \hline \hline
        $~J/\psi \, \eta_{c}~$     & $1.15\cdot10^{ 0} $ & $1.38\cdot10^{-1} $ & $9.96\cdot10^{-3} $ & $5.71\cdot10^{-4} $ & $1.04\cdot10^{-3} $ & $2.25\cdot10^{-6} $ \\ \hline 
        $~J/\psi \, J/\psi~$       & $5.67\cdot10^{-5} $ & $3.23\cdot10^{-5} $ & $1.92\cdot10^{-5} $ & $2.00\cdot10^{-5} $ & $5.70\cdot10^{-3} $ & $9.37\cdot10^{-7} $ \\ \hline
        $~\Upsilon \, \eta_{b}~$   & $-                $ & $1.40\cdot10^{-2} $ & $8.30\cdot10^{-3} $ & $2.59\cdot10^{-4} $ & $1.15\cdot10^{-3} $ & $6.47\cdot10^{-8} $ \\ \hline
        $~\Upsilon \, \Upsilon~$   & $-                $ & $9.84\cdot10^{-7} $ & $2.07\cdot10^{-5} $ & $1.17\cdot10^{-5} $ & $7.65\cdot10^{-4} $ & $3.41\cdot10^{-8} $ \\ \hline
    \end{tabular}
\end{table}

\end{document}